\documentclass[aps,twocolumn,amsmath,letterpaper,floatfix]{revtex4}
\usepackage{amssymb}
\usepackage{graphicx}
\usepackage{array}
\usepackage{hhline}
\usepackage{longtable}

\begin{document}

\title{Quantum-Mechanically Induced

Asymmetry in the Phase Diagrams of Spin-Glass Systems}
\author{C. Nadir Kaplan$^{1}$ and A. Nihat Berker$^{1-3}$}
\affiliation{$^1$Department of Physics, Ko\c{c} University, Sar\i
yer 34450, Istanbul, Turkey,} \affiliation{$^2$Feza G\"ursey
Research Institute, T\"UBITAK - Bosphorus University,
\c{C}engelk\"oy 81220, Istanbul, Turkey,}
\affiliation{$^3$Department of Physics, Massachusetts Institute of
Technology, Cambridge, Massachusetts 02139, U.S.A.}

\begin{abstract}

The spin-1/2 quantum Heisenberg spin-glass system is studied in all
spatial dimensions $d$ by renormalization-group theory.  Strongly
asymmetric phase diagrams in temperature and antiferromagnetic bond
probability $p$ are obtained in dimensions $d\geq3$.  The asymmetry
at high temperatures approaching the pure ferromagnetic and
antiferromagnetic systems disappears as $d$ is increased.  However,
the asymmetry at low but finite temperatures remains in all
dimensions, with the antiferromagnetic phase receding to the
ferromagnetic phase.  A finite-temperature second-order phase
boundary directly between the ferromagnetic and antiferromagnetic
phases occurs in $d\geq6$, resulting in a new multicritical point at
its meeting with the boundaries to the paramagnetic phase.  In
$d=3,4,5$, a paramagnetic phase reaching zero temperature intervenes
asymmetrically between the ferromagnetic and reentrant
antiferromagnetic phases.  There is no spin-glass phase in any
dimension.


\end{abstract}
\maketitle
\def\s{\rule{0in}{0.28in}}

\begin{figure*}[!t]
\centering
\includegraphics*[scale=0.99]{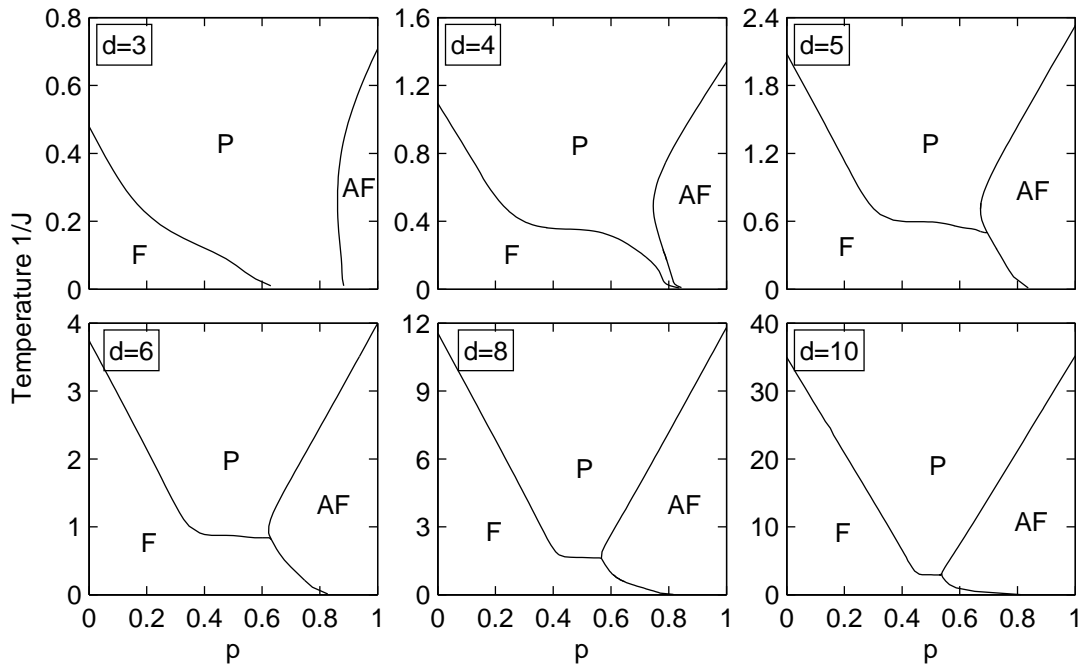}
\caption{Phase diagrams of the quantum Heisenberg spin-glass systems
in temperature $1/J$ versus antiferromagnetic bond concentration $p$
for $d=3$ to $10$.   All transitions are second-order, between the
ferromagnetic (F), antiferromagnetic(AF), and paramagnetic (P)
phases.}\label{phasediagrams}
\end{figure*}

A conspicuous \underline{finite-temperature} effect of quantum
mechanics is the critical temperature differentiation between
ferromagnetic and antiferromagnetic
systems.\cite{Rushbrooke,Oitmaa,FalicovBerker,twosuperconductingphases}
This is a contrast to classical systems where, \textit{e.g.}, on
loose-packed lattices ferromagnetic and antiferromagnetic systems
are mapped onto each other and therefore have the same critical
temperature.  We find that this quantum effect is compounded and
even more robust in spin-glass systems, which incorporate the
passage from ferromagnetism and antiferromagnetism via quenched
disorder.

Thus, in the present work, the phase diagrams of the spin-1/2
quantum Heisenberg spin-glass systems are calculated in all
dimensions $d\geq3$.  In the space of temperature $T$ and
concentration $p$ of antiferromagnetic bonds, remarkably asymmetric
phase diagrams are obtained, in very strong contrast to the
corresponding classical systems.  Whereas, in the limit of
$d\rightarrow \infty$, the differentiation of the critical
temperatures of the ferromagnetic and antiferromagnetic pure systems
disappears, the $Tp\,$ phase diagrams remain strongly asymmetric at
low but finite temperatures, where quantum fluctuations remain
dominant independent of dimensionality.  A direct second-order phase
boundary between ferromagnetic and antiferromagnetic phases, also
not seen in isotropic classical systems, is found in $d>5$. In lower
$d$, a paramagnetic phase intervenes between the ferromagnetic and
antiferromagnetic systems. Our calculation is an approximation for
hypercubic lattices and, simultaneously, a lesser approximation for
hierarchical lattices
\cite{BerkerOstlund,Kaufman,Kaufman2,Erbas,Hinczewski2,Hinczewski3,ZRZ,ZZZ,Rozenfeld,ben-Avraham,Khajeh}.

The spin-1/2 quantum Heisenberg spin-glass systems have the
Hamiltonian $-\beta \mathcal{H} = \sum_{\langle ij \rangle}
J_{ij}\mathbf{s}_i\cdot\mathbf{s}_j  \equiv \sum_{\langle ij
\rangle}-\beta \mathcal{H}(i,j)$, where $\langle ij \rangle$ denotes
a sum over pairs of nearest-neighbor sites. $J_{ij}$ is equal to the
ferromagnetic value of $J>0$ with probability $1-p$ and to the
antiferromagnetic value of $-J<0$ with probability $p$.  We solve
this model by extending the Suzuki-Takano rescaling \cite{SuzTak,
TakSuz, FalicovBerker, anisotropictJ, randomtJ,
twosuperconductingphases, KBH, Tomczak, TomRich1, TomRich2,Ozan} to
non-uniform systems and to length-rescaling factor $b=3$, necessary
for the \textit{a priori} equivalent treatment of ferromagnetism and
antiferromagnetism, followed by the essentially exact treatment
\cite{Falicov2,Migliorini} of the quenched randomness giving the
non-uniformity. In one dimension,

\begin{equation}\label{eq:3}
\begin{split} \text{Tr}_{\left(j, k\right)}e^{-\beta
\mathcal{H}}
=\text{Tr}_{\left(j, k\right)} e^{\sum_{i}^{\text{4n}}\left\{
-\beta \mathcal{H}(i,j)-\beta \mathcal{H}(j,k)-\beta \mathcal{H}(k, l) \right\}}\\
\simeq\prod_{i}^{\text{4n}}\text{tr}_{\left(j, k\right)}e^{\left\{
-\beta \mathcal{H}(i,j)-\beta \mathcal{H}(j,k)-\beta \mathcal{H}(k,
l)\right\} }
\\=\prod_{i}^{\text{
4n}}e^{-\beta ^{\prime }\mathcal{H}^{\prime }(i,l)} \simeq
e^{\sum_{i}^{\text{4n}}\left\{ -\beta ^{\prime }\mathcal{H}^{\prime
}(i,l)\right\} } =e^{-\beta ^{\prime }\mathcal{H}^{\prime }},
\end{split}
\end{equation}
where the sums and products $i$ are over every fourth spin along the
chain, the traces are over all other spins, and
$-\beta'\mathcal{H}'$ is the renormalized Hamiltonian.  Thus, the
commutation rules are correctly accounted for within four-site
segments, at all successive length scales in the iterations of the
renormalization-group transformation.  The trace tr is performed by
quantum algebra, as given below.

The rescaling is extended to dimensions $d>1$ by bond-moving, namely
by adding $b^{d-1}$ interactions resulting from the decimation of
Eq.(\ref{eq:3}), to obtain the renormalized interaction strength
$J'_{i'j'}=R(\{J_{ij}\})$, where $\{J_{ij}\}$ includes $b^d$
interactions of the unrenormalized system. The interaction constant
values $\{J_{ij}\}$ are distributed with a quenched probability
distribution ${\cal P}(J_{ij})$ \cite{Falicov2,Migliorini}, which
starts out as a double-delta function but quickly becomes
complicated under its renormalization-group transformation, given by
the convolution ${\cal P}^\prime(J^\prime_{i^\prime j^\prime}) =
\int [\prod_{ij}^{i^\prime j^\prime} dJ_{ij}\, {\cal P}(J_{ij})]
\delta(J^\prime_{i^\prime j^\prime} - R(\{J_{ij}\}))$.  This
equation actually involves $b^d$ convolutions (for example, 729
convolutions for the $d=6$ system discussed below), which are
constituted of triplet convolutions of interactions in series
(decimation) and pairwise convolutions of interactions in parallel
(bond-moving). The quenched probability distribution ${\cal
P}(J_{ij})$ is kept numerically in terms of histograms.  The number
of histograms multiplicatively increases under rescaling, until a
calculationally acceptable maximum is reached.  After this point,
the number of histograms is kept constant by implementing a binning
procedure before each pairwise or triplet convolution.  We employ a
new binning procedure, in which bins are demarked so as to contain
equal probabilities, as opposed to equal interaction intervals as
done previously.  Starting from the lowest $J$ value and moving to
greater ones, histograms in each consecutive bin are combined, to
interaction value $J = \Sigma p_iJ_i / \Sigma p_i$ and imposed equal
probability $p = \Sigma p_i = 1/n_{\text{bin}}$. In this process,
histograms at the boundaries of bins are apportioned between the
consecutive bins.  Thus, our calculation has 125,000 histograms
after each decimation and 40,000 histograms after each pairwise bond
moving.  The global flows of the quenched probability distributions
yield the phase diagrams. Analysis of the unstable fixed points and
unstable fixed distributions attracting the phase boundaries yields
the order of the phase transitions.

Calculations are done for the quantum Heisenberg spin-glass systems
in integer dimensions.  No finite-temperature phase transition
occurs in $d=1,2$.  The phase diagrams for $d=3,4,5,6,8,10$ are
shown in Fig.\ref{phasediagrams}.  They are all strikingly
asymmetric, especially in the middle $p$ and low-temperature
(would-be spin-glass phase) region.  In $d=3$, our calculated ratio
of the critical temperatures of the pure antiferromagnetic and
ferromagnetic systems is $T_C^{AF}/T_C^F=1.48$.  This value is to be
compared with the values of 1.13 found in the cubic lattice
\cite{Rushbrooke, Oitmaa} and 1.22 found in the $b=2, d=3$
hierarchical lattice \cite{FalicovBerker,twosuperconductingphases}.
This critical temperature difference is consistent with the lower
ground-state energy of the antiferromagnetic system, as calculated
\cite{Nishimori} in $d=3$.  Our calculated ratios of the
antiferromagnetic and ferromagnetic critical temperatures, for
$d=4,5,6,8,10$, decrease as 1.22, 1.12, 1.07, 1.02, 1.01
respectively. On the other hand, it is seen that although the phase
boundaries leading to the pure ferromagnetic and antiferromagnetic
critical points regain symmetry as $d$ is increased, the
low-temperature phase diagrams remain asymmetric. The ferromagnetic
phase penetrates the antiferromagnetic region at low temperatures.
Thus, quantum fluctuations present at low temperatures favor the
ferromagnetic phase over the antiferromagnetic phase. In $d\geq6$, a
second-order phase boundary occurs directly between the
ferromagnetic and antiferromagnetic phases, as is not seen in
isotropic classical spin-glass systems.  A new multicritical point
occurs where all three second-order boundaries meet.

The phase transition, between ordered phases, that is driven by
quenched randomness presents a contrast to phase transitions between
ordered phases driven by a systemwise uniform interaction. The
latter phase transition is obtained, at low temperatures, by driving
a uniform interaction that favors another ordered phase over the
existing one.  Under these conditions, essentially the entire system
remains in one ordered phase until the phase transition point is
reached, when essentially the entire system changes over to the
other phase.  Throughout this process, the ordered domains are
compact and have fractal dimensionality equal to spatial
dimensionality, which translates to having a renormalization-group
eigenvalue exponent of $y = d$, the condition for a first-order
transition \cite{Ostlund}.  By contrast, the phase transition with
quenched randomness is obtained, at low temperatures, by driving
quenched local interactions that favor the other ordered phase over
the existing phase.  This means, for example, increasing the number
of random antiferromagnetic bonds when the system is in the
ferromagnetic phase.  Under these conditions, the ferromagnetic
domains avoid the random localities of antiferromagnetic bonds in
the system.  The ferromagnetically ordered domains loose weight as
the transition is approached, so that the average magnetization
decreases.  At the phase transition, the ordered domains are not
compact and have fractal dimensionality less than the spatial
dimensionality, so that the magnetization is zero.  This translates
to the renormalization-group eigenvalue exponent $y < d$, meaning a
second-order phase transition.  The converse happens when the phase
transition is approached from the opposite side, with non-compact
antiferromagnetic domains avoiding the random localities of the
ferromagnetic bonds.  Similarly, in another recently studied system
with quenched randomness, second-order transitions between
ferromagnetic and layered ordered phases and between
antiferromagnetic and columnar ordered phases, are obtained in the
exact solution of classical anisotropic spin-glasses on a
hierarchical lattice.\cite{Guven}  Thus, we find that whereas phase
transitions between ordered phases are first order when driven by a
uniform interaction, they are second order when driven by quenched
randomness.

In $d=3,4,5$, the paramagnetic phase reaching zero temperature (as
an extremely narrow sliver in $d=5$) intervenes between the
ferromagnetic and antiferromagnetic phases. In all cases, the
ferromagnetic phase penetrates, reaching the high $p$ values of
$0.63$ and $0.83$ respectively in $d=3$, where there is a
zero-temperature paramagnetic interval, and $d\geq4$, where there is
no zero-temperature paramagnetic interval. The antiferromagnetic
phase recedes at low temperatures, thereby showing a reentrant phase
boundary \cite{Migliorini}.

There is no spin-glass phase, in the quantum system, in any
dimension.  The quantum version of the Sherrington-Kirkpatrick model
\cite{SK}, namely the spin-1/2 quantum Heisenberg model with
equivalent-neighbor interactions, with a symmetric gaussian
distribution, studied from the high-temperature side, yields a
finite-temperature phase transition, which has been interpreted as a
transition to a low-temperature spin-glass phase \cite{BM}. This
model should be similar to our studied models at $p=0.5$ in the
large $d$ limit.  Thus, we also find a finite-temperature phase
transition (Fig.1), but the low-temperature phase is explicitly a
ferromagnetic phase with quenched bond randomness.  The latter phase
has considerable amount of short-range antiferromagnetic
correlations, as seen in Ref.\cite{Burcu}.

\noindent\textit{Acknowledgments -} We are grateful to M. Hinczewski
and H. Nishimori for useful comments.  This research was supported
by the Scientific and Technical Research Council (T\"UB\.ITAK) and
by the Academy of Sciences of Turkey.


\begin{table}[h]
\begin{tabular}{|c|c|c|c|}
 \hline
  $p$ & $s$ & $m_s$ & Two-site eigenstates\\
  \hline
  $+$ & $1$ & $1$ &
  $|\phi_{1}\rangle=|\uparrow\uparrow\rangle$\\ \hline
  $+$ & $1$ & $0$ &
  $|\phi_{2}\rangle=\frac{1}{\sqrt{2}}\{|\uparrow\downarrow\rangle+|\downarrow
  \uparrow\rangle\}$\\ \hline
  $-$ & $0$ & $0$ & $|\phi_{4}\rangle=\frac{1}{\sqrt{2}}\{|\uparrow\downarrow\rangle
  -|\downarrow\uparrow\rangle\}$\\ \hline
\end{tabular}
\caption{The two-site basis states, with the corresponding parity
($p$), total spin ($s$), and total spin $z$-component ($m_s$)
quantum numbers.  The state $|\phi_{3}\rangle$ is obtained by spin
reversal from $|\phi_{1}\rangle$.  The renormalized two-site
Hamiltonian $-\beta^\prime H^\prime(i,l)$ is diagonal in this set,
with the diagonal elements of $|\phi_{1-3}\rangle$ and
$|\phi_{4}\rangle$ being $\frac{1}{4}J^\prime+G^\prime$ and
$-\frac{3}{4}J^\prime+G^\prime$ respectively.}
\end{table}

\squeezetable
\begin{table}
\begin{tabular}{|c|c|c|c|}
 \hline
  $p$ & $s$ & $m_s$ & Four-site eigenstates\\
  \hline
  $+$ & $2$ & $2$ &
  $|\psi_{1}\rangle=|\uparrow\uparrow\uparrow\uparrow\rangle$\\
  \hline
  $+$ & $2$ & $1$ &
  $|\psi_{2}\rangle=\frac{1}{2}\{|\uparrow\uparrow\uparrow\downarrow\rangle+|\uparrow\uparrow\downarrow\uparrow\rangle
  +|\uparrow\downarrow\uparrow\uparrow\rangle+|\downarrow\uparrow\uparrow\uparrow\rangle\}$\\
  \hline
  $+$ & $2$ & $0$ &
  $|\psi_{3}\rangle=\frac{1}{\sqrt{6}}\{|\uparrow\uparrow\downarrow\downarrow\rangle+|\uparrow\downarrow\uparrow\downarrow\rangle
  +|\uparrow\downarrow\downarrow\uparrow\rangle$\\
  &&&$+|\downarrow\uparrow\uparrow\downarrow\rangle+|\downarrow\uparrow\downarrow\uparrow\rangle
  +|\downarrow\downarrow\uparrow\uparrow\rangle\}$\\
  \hline
  $+$ & $1$ & $1$ &
  $|\psi_{6}\rangle=\frac{1}{2}\{|\uparrow\uparrow\uparrow\downarrow\rangle-|\uparrow\uparrow\downarrow\uparrow\rangle-|\uparrow\downarrow\uparrow\uparrow\rangle
  +|\downarrow\uparrow\uparrow\uparrow\rangle\}$\\
  \hline
  $+$ & $1$ & $0$ &
  $|\psi_{7}\rangle=\frac{1}{\sqrt{2}}\{|\downarrow\uparrow\uparrow\downarrow\rangle-|\uparrow\downarrow\downarrow\uparrow\rangle$\\
  \hline
  $-$ & $1$ & $1$ &
  $|\psi_{9}\rangle=\frac{1}{2}\{|\uparrow\uparrow\uparrow\downarrow\rangle-|\uparrow\uparrow\downarrow\uparrow\rangle+|\uparrow\downarrow\uparrow\uparrow\rangle
  -|\downarrow\uparrow\uparrow\uparrow\rangle\}$\\
  &&&$|\psi_{10}\rangle=\frac{1}{2}\{|\uparrow\uparrow\uparrow\downarrow\rangle+|\uparrow\uparrow\downarrow\uparrow\rangle-|\uparrow\downarrow\uparrow\uparrow\rangle
  -|\downarrow\uparrow\uparrow\uparrow\rangle\}$\\
  \hline
  $-$ & $1$ & $0$ &
  $|\psi_{11}\rangle=\frac{1}{\sqrt{2}}\{|\uparrow\downarrow\uparrow\downarrow\rangle-|\downarrow\uparrow\downarrow\uparrow\rangle$\\
  &&&$|\psi_{12}\rangle=\frac{1}{\sqrt{2}}\{|\uparrow\uparrow\downarrow\downarrow\rangle-|\downarrow\downarrow\uparrow\uparrow\rangle$\\
  \hline
  $+$ & $0$ & $0$ &
  $|\psi_{15}\rangle=\frac{1}{2}\{|\uparrow\uparrow\downarrow\downarrow\rangle-|\uparrow\downarrow\uparrow\downarrow\rangle-|\downarrow\uparrow\downarrow\uparrow\rangle
  +|\downarrow\downarrow\uparrow\uparrow\rangle\}$\\
  &&&$|\psi_{16}\rangle=\frac{1}{\sqrt{12}}\{|\uparrow\uparrow\downarrow\downarrow\rangle+|\uparrow\downarrow\uparrow\downarrow\rangle
  -2|\uparrow\downarrow\downarrow\uparrow\rangle$\\
  &&&$-2|\downarrow\uparrow\uparrow\downarrow\rangle+|\downarrow\uparrow\downarrow\uparrow\rangle
  +|\downarrow\downarrow\uparrow\uparrow\rangle\}$\\
  \hline
\end{tabular}
\caption {The four-site basis states, with the corresponding parity
($p$), total spin ($s$), and total spin $z$-component ($m_s$)
quantum numbers. The states $|\psi_{4,5}\rangle$,
$|\psi_{8}\rangle$, $|\psi_{13,14}\rangle$ are obtained by spin
reversal from $|\psi_{2,1}\rangle$, $|\psi_{6}\rangle$,
$|\psi_{9,10}\rangle$, respectively.}
\end{table}

\begingroup
\squeezetable
\begin{table}
\begin{gather*}
\begin{array}{|c||c|c|c|}\hline
 & \psi_{1} &\psi_{2} &\psi_{3}\\
\hhline{|=#=|=|=|} \psi_{1} &\frac{1}{4}(J_1+J_2+J_3) & 0 & 0\\
\hline \psi_{2} &0& \frac{1}{4}(J_1+J_2+J_3)&0  \\
\hline \psi_{3} &0&0 & \frac{1}{4}(J_1+J_2+J_3)\\ \hline
\end{array}\\
\begin{array}{|c||c|c|c|}\hline
 & \psi_{6} &\psi_{9} &\psi_{10}\\
\hhline{|=#=|=|=|} \psi_{6} &\frac{1}{4}(-J_1+J_2-J_3) & \frac{1}{2}(J_1-J_3) & 0\\
\hline \psi_{9} &\frac{1}{2}(J_1-J_3)& -\frac{1}{4}(J_1+J_2+J_3)&\frac{1}{2}J_2  \\
\hline \psi_{10} &0&\frac{1}{2}J_2& \frac{1}{4}(J_1-J_2+J_3)
\\ \hline
\end{array}\\
\begin{array}{|c||c|c|c|}\hline
 & \psi_{7} &\psi_{11} &\psi_{12}\\
\hhline{|=#=|=|=|} \psi_{7} &\frac{1}{4}(-J_1+J_2-J_3) & \frac{1}{2}(J_1-J_3) & 0\\
\hline \psi_{11} &\frac{1}{2}(J_1-J_3)& -\frac{1}{4}(J_1+J_2+J_3)&\frac{1}{2}J_2  \\
\hline \psi_{12} &0&\frac{1}{2}J_2& \frac{1}{4}(J_1-J_2+J_3)
\\ \hline
\end{array}\\
\begin{array}{|c||c|c|}\hline
 & \psi_{15} &\psi_{16}\\
\hhline{|=#=|=|} \psi_{15} &-\frac{3}{4}J_2 & \frac{\sqrt{3}}{4}(J_1+J_3)\\
\hline \psi_{16} & \frac{\sqrt{3}}{4}(J_1+J_3) &
\frac{1}{4}(-2J_1+J_2-2J_3)
\\ \hline
\end{array}\\
\end{gather*}
\caption{Diagonal matrix blocks of the unrenormalized three-site
Hamiltonian $-\beta H(i,j)-\beta H(j,k)-\beta H(k,l)$.  The
Hamiltonian being invariant under spin-reversal, the spin-flipped
matrix elements are not shown.  The additive constant $3G$ at the
diagonal elements is not shown.  The interaction constants
$J_1,J_2,J_3$, which are in general unequal due to quenched
randomness, are from $-\beta \mathcal{H}(i,j),-\beta
\mathcal{H}(j,k),-\beta \mathcal{H}(k,l)$ respectively.}
\end{table}
\endgroup


\noindent\textit{Appendix -} The operators $-\beta^\prime
\mathcal{H}^\prime(i, l)$ and $-\beta \mathcal{H}(i, j)-\beta
\mathcal{H}(j, k)-\beta \mathcal{H}(k, l)$ of Eq. \eqref{eq:3}  act
on two-site and four-site states, respectively, where at each site
the spin is in quantum state $\sigma=\uparrow$ or $\downarrow$.  The
trace tr in Eq. \eqref{eq:3} is, in terms of matrix elements
\cite{FalicovBerker},

\begin{multline}
\langle u_{i}z_{l}|e^{-\beta ^{\prime }\mathcal{H}^{\prime
}(i,l)}|\bar{u}_{i}^{{}}\bar{z}_{l}^{{}}\rangle = \label{eq:app1}\\
\sum_{v_{j}, w_{k}}\langle u_{i}\,v_{j}\,w_{k}z_{l}|e^{-\beta
\mathcal{H}(i,j)-\beta \mathcal{H}(j,k)-\beta
\mathcal{H}(k,l)}|\bar{u}_{i}\,v_{j}\,w_{k}\bar{z}_{l}^{{}}\rangle
\:,
\end{multline}
where $u_{i},v_{j},w_{k}, z_{l},\bar{u}_{i},\bar{z}_{l}^{{}}$ are
single-site state variables.  Thus, Eq. \eqref{eq:app1} is the
contraction of a $16\times16$ matrix into a $4\times4$ matrix. Basis
states that are simultaneous eigenstates of parity ($p$), total spin
magnitude ($s$), and total spin z-component ($m_s$)
block-diagonalize these matrices and thereby make Eq.
\eqref{eq:app1} manageable.  These sets of 4 two-site and 16
four-site eigenstates, denoted by $\{|\phi _{p}\rangle \}$ and
$\{|\psi _{q}\rangle \}$ respectively, are given in Tables I and II.
The diagonal blocks are given in Tables I and III.  Due to the
microscopic randomness of the spin-glass problem, the four-site
Hamiltonian mixes states of different parity, as seen in Table IV.
Eq. \eqref{eq:app1} is thus rewritten as

\begin{multline}\label{eq:app2}
\langle \phi _{p}|e^{-\beta ^{\prime }\mathcal{H}^{\prime }(i,k)}|\phi _{\bar{p}%
}\rangle = \sum_{\substack{u,z,\bar{u},\\ \bar{z},v, w}}
\sum_{\substack{q,\bar{q}}} \langle\phi _p|u_iz_l\rangle \langle
u_iv_jw_kz_l|\psi_q\rangle\\ \langle \psi _q|e^{-\beta
\mathcal{H}(i,j)-\beta \mathcal{H}(j,k)-\beta \mathcal{H}(k,l)}|\psi
_{\bar{q}}\rangle\langle
\psi_{\bar{q}}|\bar{u}_iv_jw_k\bar{z}_l\rangle \langle
\bar{u}_i\bar{z}_l|\phi _{\bar{p}}\rangle\:.
\end{multline}
There are only two rotation-symmetry independent elements of
$\langle \phi _{p}|e^{-\beta ^{\prime }\mathcal{H}^{\prime
}(i,l)}|\phi_{\bar{p}}\rangle \equiv
\langle\phi_{p}||\phi_{\bar{p}}\rangle$ in Eq.\eqref{eq:app2}, which
have $p=\bar{p}=1,4$ (thereby leading to one renormalized
interaction constant $J'$ and the additive constant $G'$).  From Eq.
\eqref{eq:app2},
$\langle\phi_{1}||\phi_{1}\rangle=\langle\psi_{1}||\psi_{1}\rangle+\frac{1}{2}\langle\psi_{2}||\psi_{2}\rangle+
\frac{1}{6}\langle\psi_{3}||\psi_{3}\rangle+
\frac{1}{2}\langle\psi_{6}||\psi_{6}\rangle+\frac{1}{2}\langle\psi_{7}||\psi_{7}\rangle
+\frac{1}{2}\langle\psi_{9}||\psi_{9}\rangle-\langle\psi_{9}||\psi_{10}\rangle+
\frac{1}{2}\langle\psi_{10}||\psi_{10}\rangle+\frac{1}{3}\langle\psi_{16}||\psi_{16}\rangle$
and
$\langle\phi_{4}||\phi_{4}\rangle=\langle\psi_{9}||\psi_{9}\rangle+2\langle\psi_{9}||\psi_{10}\rangle+\langle\psi_{10}||\psi_{10}\rangle+
\frac{1}{2}\langle\psi_{11}||\psi_{11}\rangle+\langle\psi_{11}||\psi_{12}\rangle+\frac{1}{2}\langle\psi_{12}||\psi_{12}\rangle+\langle\psi_{15}||\psi_{15}\rangle$,
with $\langle\psi_{q}||\psi_{\bar{q}}\rangle\equiv\langle \psi
_{q}|e^{-\beta \mathcal{H}(i,j)-\beta \mathcal{H}(j,k)-\beta
\mathcal{H}(k,l)}|\psi_{\bar{q}}\rangle$. From Table I, the
renormalized interaction constant is given by
$J'=\ln({\langle\phi_{1}||\phi_{1}\rangle/\langle\phi_{4}||\phi_{4}\rangle})$.

\end{document}